\documentclass[aip,jcp,reprint,amsmath,amssymb,floatfix,citeautoscript,nofootinbib]{revtex4-1}
\usepackage{times}
\usepackage{mathptmx}
\DeclareMathAlphabet{\mathcal}{OMS}{cmsy}{m}{n} 
\usepackage[T1]{fontenc}
\usepackage{algorithm}
\usepackage{algpseudocode}
\usepackage{bm}
\usepackage{amsmath}
\usepackage{amssymb}
\usepackage{xcolor}
\usepackage{graphicx}
\usepackage{hyperref}
\hypersetup{
    colorlinks=true,
    linkcolor=blue,
    urlcolor=blue,
    citecolor=blue,
}
\usepackage{cleveref}
\usepackage{braket}
\usepackage{mhchem}
\usepackage{multirow}

\newcommand{\mi}{\mathrm{i}}

\definecolor{myred}{rgb}{0.8,0,0}

\begin{document}

    \title{Efficient Implementation of the Random Phase Approximation with Domain-based Local Pair Natural Orbitals}

    \author{Yu Hsuan Liang}
    \affiliation{Department of Chemistry, Columbia University, New York, NY, 10027, USA}
    \author{Xing Zhang}
    \affiliation{Division of Chemistry and Chemical Engineering, California Institute of Technology, Pasadena, California 91125, USA}
    \author{Garnet Kin-Lic Chan}
    \affiliation{Division of Chemistry and Chemical Engineering, California Institute of Technology, Pasadena, California 91125, USA}
    \author{Timothy C. Berkelbach}
    \email{tim.berkelbach@gmail.com}
    \affiliation{Department of Chemistry, Columbia University, New York, NY, 10027, USA}
    \affiliation{Initiative for Computational Catalysis, Flatiron Institute, New York, NY, 10010, USA}
    \author{Hong-Zhou Ye}
    \email{hzye@umd.edu}
    \affiliation{Department of Chemistry and Biochemistry, University of Maryland, College Park, MD, 20742, USA}
    \affiliation{Institute for Physical Science and Technology, University of Maryland, College Park, MD, 20742, USA}

    \begin{abstract}
        We present an efficient implementation of the random phase approximation (RPA) for molecular systems within the domain-based local pair natural orbital (DLPNO) framework. With optimized parameters, DLPNO-RPA achieves approximately $99.9\%$ accuracy in the total correlation energy compared to a canonical implementation, enabling highly accurate reaction energies and potential energy surfaces to be computed while substantially reducing computational costs.
        As an application, we demonstrate the capability of DLPNO-RPA to efficiently calculate basis set-converged binding energies for a set of large molecules, with results showing excellent agreement with high-level reference data from both coupled cluster and diffusion Monte Carlo.
        This development paves the way for the routine use of RPA-based methods in molecular quantum chemistry.
    \end{abstract}

    \maketitle

    \section{Introduction}

    Positioned on the fifth rung of Jacob's ladder~\cite{Perdew01AIPCP}, the random phase approximation~\cite{Bohm53PR,GellMann57PR,Ren12JMS,Zhang21WIRCMS} (RPA) is a powerful approach for incorporating non-local many-electron correlation effects into Kohn-Sham density functional theory~\cite{Hohenberg64PR,Kohn65PR} (DFT).
    RPA's strength in capturing long-range dispersion interactions and the accurate modeling of metallic systems has made it a valuable tool for applications such as surface adsorption~\cite{Schimka10NM,Ren09PRB,Olsen11PRL,Rohlfin08PRL,Ma11PRB,Kim12PRB,Schmidt18JPCC,Sheldon24JCTC},
    molecular crystals~\cite{Lu09PRL,DelBen13JCTC,Klimes16JCP,Zen18PNAS,Pham23JCP}, layered materials~\cite{Marini06PRL,Hellgren21PRR},
    liquid water~\cite{Yao21JPCL,Villard24CS},
    and thermochemistry~\cite{Ren13PRB,Waitt16JCTC,Bates17JCTC,Chedid18TCA,Millican22CM,Rey24CST}.
    Despite these successes, routine RPA calculations for systems of more than 100 atoms in the complete basis set (CBS) limit remain computationally challenging.

    Most practical implementations of RPA rely on the adiabatic connection fluctuation-dissipation theorem (ACFDT)~\cite{Gunnarsson76PRB,Langreth77PRB}, which has a computational cost scaling of $\mathcal{O}(N^4)$ with system size $N$, using either plane-wave basis sets or atom-centered basis sets with the density fitting (DF) approximation [also known as resolution of the identity (RI)]~\cite{Ren12NJP,Eshuis10JCP,DelBen13JCTC}.
    Recent developments, including the space-time algorithm~\cite{Rojas95PRL,Kaltak14PRB}, local RI~\cite{Shi24PRB} or RI with an overlap metric~\cite{Wilhelm16JCTC}, stochastic sampling~\cite{Neuhauser13JPCL}, and interpolative separable DF~\cite{Lu17JCP}, have been demonstrated to reduce this scaling and therefore increase its application range for materials simulations.
    Alternatively, RPA can be formulated within the coupled-cluster~\cite{Bartlett07RMP} (CC) framework as direct ring CC with double excitations (drCCD)~\cite{Scuseria08JCP}---an approximation to CC with single and double excitations (CCSD) that retains only the direct particle-hole ring diagrams.
    Although CC-based RPA typically scales as $\mathcal{O}(N^6)$ like CCSD, the cost can be reduced to $\mathcal{O}(N^4)$ using DF techniques~\cite{Scuseria08JCP,Hebelmann12PRA,Kallay15JCP}.
    Within this framework, a linear-scaling implementation of RPA has been reported using the local natural orbital (LNO) approximation~\cite{Kallay15JCP}.

    In this work, we present a reduced-scaling implementation of CC-based RPA using the domain-based local pair natural orbital (DLPNO) approximation~\cite{Neese09JCPa,Neese09JCPb}.
    Compared to ACFDT-RPA or LNO-RPA, an advantage of the DLPNO-RPA method is that it yields the global wavefunction amplitudes, in addition to the correlation energy, which facilitates calculations of other properties.
    As a versatile local correlation framework, DLPNO has been successfully applied to achieve linear scaling in methods such as second-order M{\o}ller-Plesset perturbation theory~\cite{Pinski15JCP,Werner15JCTC} (MP2), CCSD~\cite{Neese09JCPa,Riplinger13JCPa}, and its extension CCSD(T)~\cite{Riplinger13JCPb,Riplinger16JCP}, which includes perturbative triple excitations~\cite{Raghavachari89CPL}.
    The accuracy of DLPNO-based methods can be finely controlled by adjusting PNO truncation thresholds, often achieving better than $99.7\%$ accuracy in the total correlation energy and deviations of less than $1$~kcal/mol in energy differences compared to the corresponding canonical methods~\cite{Liakos15JCTC}.
    In the following, we detail the implementation of DLPNO-RPA, highlighting its high accuracy and computational efficiency compared to canonical ACFDT-RPA across a range of numerical examples.
    As an application, we calculate basis set-converged binding energies for a series of large molecules using DLPNO-RPA with a PBE reference and find excellent agreement with high-level correlated wavefunction methods at only a fraction of the computational cost.

    \section{Theory}

    Our implementation of DLPNO-RPA closely follows previous work by Werner and co-workers on PNO-MP2~\cite{Werner15JCTC} (especially for domain construction) and work by Neese and co-workers on DLPNO-MP2~\cite{Pinski15JCP} and DLPNO-CCSD~\cite{Riplinger16JCP}.
    Throughout the discussion, we assume a spin-restricted Hartree-Fock (RHF) reference whose occupied orbitals can be localized by a unitary transformation.
    We use $i,j,\ldots$ for localized occupied molecular orbitals (LMOs), $a,b,\ldots$ for canonical virtual orbitals, $r,s,\ldots$ for projected atomic orbitals~\cite{Pulay83CPL,Sabo85CPL} (PAOs), $a_{i}, b_{j}, \ldots$ for orbital-specific virtuals~\cite{Yang11JCP,Yang12JCP} (OSVs),
    $\bar{a}_{ij}, \bar{b}_{ij}, \ldots$ for joint OSVs, $a_{ij}, b_{ij}, \ldots$ for PNOs, and $P,Q,\ldots$ for auxiliary orbitals.
    Unless otherwise specificied, all virtual orbitals, including PAOs, OSVs, joint OSVs, and PNOs, are assumed to be semi-canonicalized;
    the corresponding orbital energies are denoted by $\varepsilon_{r}, \varepsilon_{a_i}$, etc.
    For four-tensors of the form $X_{iajb}$, we will interchangeably denote them by matrices $(\mathbf{X}^{ij})_{ab}$.
    Domains specific to a LMO $i$ or a LMO pair $(i,j)$ will be denoted by $[i]$ and $[ij]$.

    In DLPNO-RPA, the total correlation energy of a molecule is calculated by summing contributions from LMO pairs of three classes,
    \begin{equation}
        E_{\textrm{corr}}^{\textrm{DLPNO-RPA}}
            = \sum_{i,j}^{N_{\textrm{pair}}^{\textrm{strong}}} E_{ij}^{\textrm{RPA}}
            + \sum_{i,j}^{N_{\textrm{pair}}^{\textrm{weak}}} E_{ij}^{\textrm{SC-dMP2}}
            + \sum_{i,j}^{N_{\textrm{pair}}^{\textrm{dist}}} E_{ij}^{\textrm{dip}}
    \end{equation}
    where strong pairs are treated fully self-consistently by solving the RPA residual equations, weak pairs are approximated at the semi-canonical direct MP2 (SC-dMP2) level, and distant pairs are approximated by a multipole expansion truncated at the dipole level.
    Different pairs are selected based on their estimated pair energy.
    First, a crude estimation of the pair energy is obtained for all pairs using the dipole approximation~\cite{Pinski15JCP,Werner15JCTC}
    \begin{equation}
        E^{\textrm{dip}}_{ij}
            = \frac{8}{R_{ij}^6} \sum_{r_i}^{[i]_{\textrm{PAO}}} \sum_{s_j}^{[j]_{\textrm{PAO}}} \frac{
                |V^{\textrm{dip}}_{i r_{i} j s_{j}}|^2
            }{
                f_{ii} + f_{jj} - \varepsilon_{r_i} - \varepsilon_{s_j}
            }
    \end{equation}
    where the summation runs over the PAO domains of LMOs $i$ and $j$, $R_{ij}$ is the distance between the centroids of the two LMOs, and the dipole approximated ERIs are~\cite{Pinski15JCP}
    \begin{equation}
        V^{\textrm{dip}}_{i r_{i} j s_{j}}
            = \bm{d}_{ir_i} \cdot \bm{d}_{js_j} - 3 (\bm{e}_{ij} \cdot \bm{d}_{ir_i}) (\bm{e}_{ij} \cdot \bm{d}_{js_j})
    \end{equation}
    where $\bm{d}_{i r_i} = \braket{i| \hat{\bm{r}} | r_{i}}$ ($\hat{\bm{r}}$ is relative to the centroid of the molecule) and $\bm{e}_{ij} = \bm{R}_{ij} / R_{ij}$.
    The construction of various types of domains will be detailed below.
    A cutoff $T_{\textrm{dist}}$ is then applied to select distant pairs by
    \begin{equation}
        |E^{\textrm{dip}}_{ij}| < T_{\textrm{dist}}.
    \end{equation}
    For pairs surviving this prescreening, the pair energy is refined using SC-dMP2
    \begin{equation}    \label{eq:eij_scmp2_osv}
        E_{ij}^{\textrm{SC-dMP2}}
            = 2 (2-\delta_{ij}) \sum_{\bar{a}_{ij} \bar{b}_{ij}}^{[ij]_{\textrm{OSV}}}
            T^{(1)}_{i \bar{a}_{ij} j \bar{b}_{ij}} V_{i \bar{a}_{ij} j \bar{b}_{ij}}
    \end{equation}
    where the summation is over the joint OSV domain of the LMO pair, $V_{i \bar{a}_{ij} j \bar{b}_{ij}} = (i \bar{a}_{ij} | j \bar{b}_{ij})$, and the (opposite-spin) SC-MP1 amplitudes are
    \begin{equation}    \label{eq:sc_mp1_amp_osv}
        T^{(1)}_{i \bar{a}_{ij} j \bar{b}_{ij}}
            = \frac{ V_{i \bar{a}_{ij} j \bar{b}_{ij}}^* }
            {f_{ii} + f_{jj} - \varepsilon_{\bar{a}_{ij}} - \varepsilon_{\bar{b}_{ij}}}
    \end{equation}
    A second cutoff $T_{\textrm{weak}}$ is applied to select weak pairs by
    \begin{equation}
        |E_{ij}^{\textrm{SC-dMP2}}|
            < T_{\textrm{weak}}
    \end{equation}
    For the remaining strong pairs, we solve the RPA amplitude equations within the PNO domains obtained by compressing the joint OSV domains for each pair
    \begin{equation}    \label{eq:rpa_amp_eqn}
    \begin{split}
        \bm{\Delta}^{ij} \circ \mathbf{T}^{ij}
            &- \sum_{k \neq i} f_{ik} \mathbf{S}^{ij,kj} \mathbf{T}^{kj} \mathbf{S}^{ij,kj\dagger}
            - \sum_{k \neq j} f_{kj} \mathbf{S}^{ij,ik} \mathbf{T}^{kj} \mathbf{S}^{ij,ik\dagger}   \\
            &+ 4 \sum_{kl} \bigg[
                \bigg(
                    \mathbf{S}^{ij,ik} \mathbf{T}^{ik} \mathbf{S}^{ik,kl}
                    + \frac{1}{2} \delta_{ik} \mathbf{S}^{ij,il}
                \bigg) \mathbf{V}^{kl}  \\
            &\phantom{+ 4 \sum_{kl} \bigg[}
            ~\bigg(
                    \mathbf{S}^{kl,lj} \mathbf{T}^{lj} \mathbf{S}^{lj,ij}
                    + \frac{1}{2} \delta_{jl} \mathbf{S}^{kj,ij}
                \bigg)
            \bigg]
            = \bm{0}
    \end{split}
    \end{equation}
    where ``$\circ$'' denotes elementwise product and $(\bm{\Delta}^{ij})_{a_{ij} b_{ij}} = f_{ii}+f_{jj}-\varepsilon_{a_{ij}}-\varepsilon_{b_{ij}}$.
    \Cref{eq:rpa_amp_eqn} is derived from the drCCD amplitude equations~\cite{Scuseria08JCP} by incorporating the PNO overlap matrices
    \begin{equation}    \label{eq:pno_ovlp}
        (\mathbf{S}^{ij,ik})_{a_{ij}b_{ik}}
            = \braket{\psi_{a_{ij}} | \psi_{b_{ik}}}
    \end{equation}
    to account for the non-orthogonality between PNOs associated with different LMO pairs.
    The $\mathbf{T}$ amplitudes in \cref{eq:rpa_amp_eqn} are not anti-symmetrized.
    The strong pair energy is given by
    \begin{equation}
        E^{\textrm{RPA}}_{ij}
            = 2 \mathrm{Tr}~\mathbf{T}^{ij} \mathbf{V}^{ij} + \Delta E_{ij}^{\textrm{PNO}}
    \end{equation}
    where $\Delta E_{ij}^{\textrm{PNO}}$ accounts for the contribution from discarded PNOs [see \cref{eq:pno_selection}], which we estimate by the difference between the SC-dMP2 energy evaluated within the joint OSV domain (\ref{eq:eij_scmp2_osv}) and the PNO domain, i.e.,
    \begin{equation}    \label{eq:pno_correction}
        \Delta E_{ij}^{\textrm{PNO}}
            = E^{\textrm{SC-dMP2}}_{ij} - E^{\textrm{PNO-SC-dMP2}}_{ij}
    \end{equation}
    where $E^{\textrm{PNO-SC-dMP2}}_{ij}$ is evaluated similarly to \cref{eq:eij_scmp2_osv}, but within the PNO domain.
    In the DLPNO-RPA framework, the second-order screened exchange~\cite{Gruneis09JCP} (SOSEX) energy can be computed with minimal additional computational cost
    \begin{equation}
        E^{\textrm{RPA+SOSEX}}_{ij}
            = \mathrm{Tr}~\mathbf{T}^{ij} (2\mathbf{V}^{ij} - \mathbf{V}^{ij\textrm{T}}) + \Delta E_{ij}^{\textrm{PNO}}
    \end{equation}
    with the additional modification that SC-dMP2, used for approximating the weak pair energy (\ref{eq:eij_scmp2_osv}) and the PNO correction (\ref{eq:pno_correction}), is now replaced by SC-MP2, i.e.,
    \begin{equation}
        E_{ij}^{\textrm{SC-MP2}}
            = (2-\delta_{ij}) \sum_{\bar{a}_{ij} \bar{b}_{ij}}^{[ij]_{\textrm{OSV}}}
            T^{(1)}_{i \bar{a}_{ij} j \bar{b}_{ij}} (2V_{i \bar{a}_{ij} j \bar{b}_{ij}} - V_{i \bar{b}_{ij} j \bar{a}_{ij}})
    \end{equation}
    and a similar expression holds for $E^{\textrm{PNO-SC-MP2}}_{ij}$.
    The SOSEX contribution has been shown to enhance the accuracy of RPA in thermochemistry applications~\cite{Gruneis09JCP,Ren13PRB} by approximately incorporating short-range exchange effects.

    The various types of domains mentioned above are determined as follows.
    First, a PAO domain is constructed for each LMO using the geometric approach proposed by Werner and co-workers~\cite{Werner15JCTC}.
    Specifically, a small subset of atoms for each LMO is first selected based on the meta-L{\"o}wdin population~\cite{Sun14JCTC} exceeding $0.2$.
    These small atom sets are expanded by including all neighboring atoms within $n_{\textrm{bond}}^{\textrm{PAO}}$ chemical bonds (two atoms are considered bonded if the distance between them is less than $1.2$ times the sum of their atomic radii) or within a distance of $d_{\textrm{bond}}^{\textrm{PAO}} = 2n_{\textrm{bond}}^{\textrm{PAO}} + 1$~Bohr.
    The PAO domain for each LMO consists of all PAOs belonging to the extended atom set corresponding to that LMO.
    Within each PAO domain, OSVs are calculated via $\psi_{a_i} = \sum_{r_i} \psi_{r_i} U_{r_i a_i}$ by diagonalizing the diagonal SC-MP1 amplitudes~\cite{Yang11JCP,Yang12JCP}
    \begin{equation}    \label{eq:osv_gen}
        (\mathbf{T}^{ii})_{r_i s_i}
            = \frac{V^*_{i r_i i s_i}}{2f_{ii} - \varepsilon_{r_i} - \varepsilon_{s_i}}
            = \sum_{a_{i}} \lambda_{a_i} U_{r_i a_i} U_{s_i a_i}^*.
    \end{equation}
    A cutoff $T_{\textrm{OSV}}$ is applied to select important OSVs that comprise the OSV domain
    \begin{equation}
        |\lambda_{a_i}| \geq T_{\textrm{OSV}}.
    \end{equation}
    The joint OSV domain for a LMO pair is constructed by combining the two sets of OSVs, followed by symmetric orthogonalization (with a cutoff of $10^{-6}$ to eliminate potential linear dependencies).
    Within each joint OSV domain, PNOs are calculated via $\psi_{a_{ij}} = \sum_{\bar{a}_{ij}} \psi_{\bar{a}_{ij}} U_{\bar{a}_{ij} a_{ij}}$ by diagonalizing the SC-MP2 pair density matrix
    \begin{equation}    \label{eq:pno_gen}
        (\mathbf{D}^{ij})_{\bar{a}_{ij} \bar{b}_{ij}}
            = \sum_{a_{ij}} \lambda_{a_{ij}} U_{\bar{a}_{ij} a_{ij}} U_{\bar{b}_{ij} a_{ij}}^*
    \end{equation}
    where
    \begin{equation}    \label{eq:sc_mp2_pairden}
        \mathbf{D}^{ij}
            = \frac{1}{1+\delta_{ij}} (
                \tilde{\mathbf{T}}^{ij\dagger} \mathbf{T}^{ij}
                + \tilde{\mathbf{T}}^{ij} \mathbf{T}^{ij\dagger}
            )
    \end{equation}
    and $\tilde{\mathbf{T}}^{ij} = 2\mathbf{T}^{ij} - \mathbf{T}^{ij\mathrm{T}}$.
    Two criteria are applied simultaneously to select important PNOs that comprise the PNO domain: an eigenvalue cutoff $T_{\textrm{PNO}}$
    \begin{equation}    \label{eq:pno_selection}
        |\lambda_{a_{ij}}| \geq T_{\textrm{PNO}}
    \end{equation}
    and an energy cutoff
    \begin{equation}    \label{eq:pno_selection_energy}
        \frac{|E_{ij}^{\textrm{PNO-SC-MP2}}|}{|E_{ij}^{\textrm{SC-MP2}}|}
            \geq T_{\textrm{EPNO}}
    \end{equation}
    As noted by Werner and co-workers~\cite{Werner15JCTC}, the energy criterion (\ref{eq:pno_selection_energy}) helps prevent PNO domains from being too small in cases where LMOs are well separated.
    We found that a small value of $T_{\textrm{EPNO}} = 0.9$ is sufficient for this purpose.

    Local DF~\cite{Werner15JCTC,Pinski15JCP} is used to efficiently evaluate the $(ov|ov)$-type ERIs needed both for generating the OSVs (\ref{eq:osv_gen}) and the PNOs (\ref{eq:pno_gen}) and for solving the amplitude equations (\ref{eq:rpa_amp_eqn}).
    First, a fitting domain $[i]_{\textrm{fit}}$ is determined for each LMO similarly to the construction of the PAO domain, with a separate parameter $n^{\textrm{fit}}_{\textrm{bond}}$ controlling its size.
    The PAO and fitting domains are both extended by taking the union of all other domains whose corresponding LMOs form a non-distant pair with the center LMO.
    These extended domains will be denoted by $[i]_{\textrm{ePAO}}$ and $[i]_{\textrm{efit}}$, respectively.
    The three-center Coulomb integrals are computed in the AO basis and then transformed into the LMO-PAO basis
    \begin{subequations}    \label{eq:irP}
    \begin{align}
        &(i\nu|P)
            = \sum_{\mu} L_{\mu i}^* (\mu\nu|P),
        \quad{} P \in [i]_{\textrm{efit}}
        \label{subeq:inuP}
        \\
        &(i r| P)
            = \sum_{\nu} O_{\nu r} (i\nu|P),
        \quad{} r \in [i]_{\textrm{ePAO}}
        \label{subeq:irP}
    \end{align}
    \end{subequations}
    where $\mathbf{L}$ and $\mathbf{O}$ are the LMO and raw PAO (i.e.,~before semi-canonicalization) coefficient matrices, respectively.
    The transformed three-center integrals are then fitted for each pair of LMOs
    \begin{subequations}    \label{eq:fit_3c_ij}
    \begin{align}
        &(i r_{ij} | \tilde{P}_{ij})
            = \sum_{Q_{ij} \in [ij]_{\textrm{fit}}} (i r_{ij} | Q_{ij}) (\mathbf{J}^{-1/2})_{Q_{ij} \tilde{P}_{ij}} \\
        &(j r_{ij} | \tilde{P}_{ij})
            = \sum_{Q_{ij} \in [ij]_{\textrm{fit}}} (j r_{ij} | Q_{ij}) (\mathbf{J}^{-1/2})_{Q_{ij} \tilde{P}_{ij}}
    \end{align}
    \end{subequations}
    where $J_{PQ} = (P|Q)$, $\tilde{P}$ denotes auxiliary index after fitting, and $[ij]_{\textrm{fit}} = [i]_{\textrm{fit}} \cup [j]_{\textrm{fit}}$.
    Finally, the PAO-basis $(ov|ov)$-type ERIs can be assembled by
    \begin{equation}    \label{eq:pao_eris}
        V_{i r_{ij} j s_{ij}}
            = \sum_{\tilde{P}_{ij}} (i r_{ij} | \tilde{P}_{ij}) (j s_{ij} | \tilde{P}_{ij})
    \end{equation}
    while ERIs in other bases are obtained by first transforming the three-center integrals and then assembling, e.g.,~for PNOs
    \begin{subequations}    \label{eq:pno_eris}
    \begin{align}
        &(i a_{ij} | \tilde{P}_{ij})
            = \sum_{r_{ij}} (i r_{ij} | \tilde{P}_{ij}) U_{r_{ij} a_{ij}}
        \\
        &V_{i a_{ij} j b_{ij}}
            = \sum_{\tilde{P}_{ij}} (i a_{ij} | \tilde{P}_{ij}) (j b_{ij} | \tilde{P}_{ij})
    \end{align}
    \end{subequations}
    where $U_{r_{ij} a_{ij}}$ transforms raw PAOs to PNOs for the LMO pair $(i,j)$.

    \section{Computational details}

    We implemented the DLPNO-RPA method outlined above in a developer version of PySCF 2.7~\cite{Sun18WIRCMS,Sun20JCP}, which uses Libcint~\cite{Sun15JCC} for AO integral evaluation.
    OpenMP is used for shared-memory multithreading.
    In the asymptotic regime, where the sizes of all domains remain independent of the system size $N$, the code achieves linear scaling in all components except for the first half-transformation (\ref{subeq:inuP}), which scales as $\mathcal{O}(N^3)$ because we do not exploit the sparsity of the LMO coefficient matrix~\cite{Pinski15JCP} in this work.
    The storage cost also shows linear scaling in the asymptotic limit, although the prefactors vary for different terms.
    The main storage bottleneck arises from the transformed three-center integrals in the extended PAO and fitting domains (\ref{subeq:irP}), which can exceed available memory and may require the use of disk space.
    All other tensors in the PNO basis needed for solving the amplitude equations (\ref{eq:rpa_amp_eqn}) can be readily stored in memory.

    In the following section (\cref{sec:results}), we benchmark the accuracy and efficiency of DLPNO-RPA against canonical ACFDT-RPA and drCCD-RPA, across a series of molecules whose geometries are provided in the Supporting Information.
    The canonical ACFDT-RPA correlation energy is calculated via
    \begin{equation}    \label{eq:acfdt_rpa_energy}
        E_{\textrm{corr}}^{\textrm{RPA}}
            = \frac{1}{2\pi} \int_0^\infty \mathrm{d}\omega~
            \textrm{Tr}\left\{
                \ln\left[
                    1 - \chi^0(\mi\omega)v
                \right] + \chi^0(\mi\omega)v
            \right\}
    \end{equation}
    with frequency integration performed using a modified Gauss-Legendre grid with $40$ points~\cite{Ren12NJP} as implemented in PySCF~\cite{Zhu21JCTC}.
    [In \cref{eq:acfdt_rpa_energy}, $\chi^0$ is the independent density-response function on imaginary frequency, $v$ is the bare Coulomb potential, $\textrm{Tr}(AB) = \int\mathrm{d}\bm{r}\mathrm{d}\bm{r}'~A(\bm{r},\bm{r}')B(\bm{r}',\bm{r})$.]
    The canonical drCCD-RPA is implemented using the DF-based quartic-scaling algorithm described in ref~\onlinecite{Kallay15JCP}.
    This implementation also enables the calculation of the canonical SOSEX energy, which we use to benchmark the performance of DLPNO-RPA+SOSEX.


    \section{Results and discussion}
    \label{sec:results}

    \begin{table}[!t]
        \centering
        \caption{Recommended parameter values for DLPNO-RPA.}
        \label{tab:param}
        \begin{tabular}{lll}
            \hline\hline
            Parameter & Explanation & Value   \\
            \hline
            $n_{\textrm{bond}}^{\textrm{PAO}}$ & PAO domain & $4$    \\
            $n_{\textrm{bond}}^{\textrm{fit}}$ & Fitting domain & $3$    \\
            $T_{\textrm{dist}}$ & Distant pairs & $10^{-6}$ \\
            $T_{\textrm{weak}}$ & Weak pairs & $3\times10^{-6}$ \\
            $T_{\textrm{OSV}}$ & OSV truncation & $10^{-4}$  \\
            $T_{\textrm{PNO}}$ & PNO truncation & $3 \times 10^{-7}$  \\
            \hline
        \end{tabular}
    \end{table}

    \begin{figure}[!b]
        \centering
        \includegraphics[width=8cm]{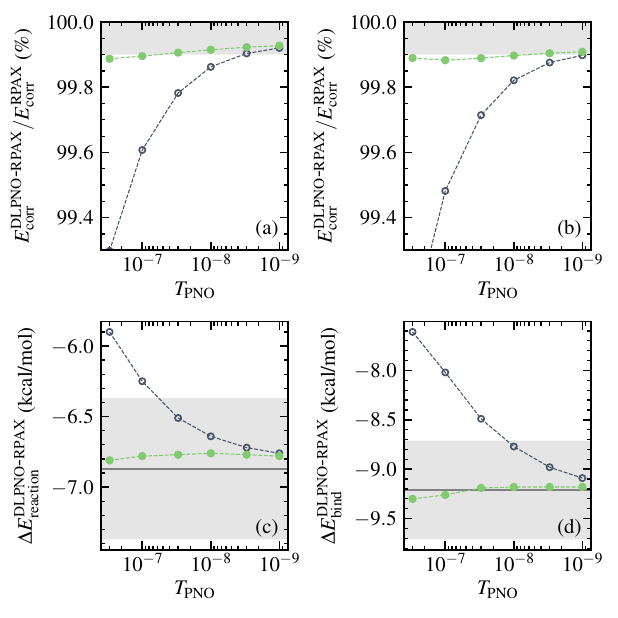}
        \caption{(a,b) Convergence of the DLPNO-RPA correlation energy as a function of $T_{\textrm{PNO}}$ for (a) an androstenedione precursor and (b) a coronene dimer.
        The results are presented both with (filled symbols) and without (open symbols) applying the PNO correction (\ref{eq:pno_correction}).
        The gray shaded area indicates better than $99.9\%$ accuracy compared to canonical ACFDT-RPA.
        (c,d) Convergence of the DLPNO-RPA energy difference as a function of $T_{\textrm{PNO}}$ for (c) the reaction energy for forming androstenedione and pyrocatechol from the precursor and (d) the binding energy of the coronene dimer.
        The gray shaded area indicates less than $\pm 0.5$~kcal/mol error compared to canonical ACFDT-RPA.
        All calculations were performed using the cc-pVTZ/cc-pVTZ-RI basis sets with core electrons being frozen.}
        \label{fig:pno_conv}
    \end{figure}

    In \cref{tab:param}, we list our recommended values for the various numerical parameters used in DLPNO-RPA.
    These values, consistent with those used in previous studies~\cite{Werner15JCTC,Liakos15JCTC}, are set relatively tight and have been shown to achieve approximately $99.9\%$ accuracy in the total correlation energy, with typical errors below $0.5$~kcal/mol in reaction energies.
    To assess convergence, we compare DLPNO-RPA results to canonical ACFDT-RPA results.
    Two examples are shown in \cref{fig:pno_conv}, illustrating convergence behavior for both covalent and weakly bound molecules.
    In \cref{fig:pno_conv}a, we show the convergence of the DLPNO-RPA correlation energy for an androstenedione precursor molecule as a function of $T_{\textrm{PNO}}$, with all other parameters set to the defaults specified in \cref{tab:param}.
    A corresponding plot for the coronene dimer is presented in \cref{fig:pno_conv}b.
    In both cases, $99.9\%$ of the RPA correlation energy is recovered for $T_{\textrm{PNO}} \leq 3\times10^{-7}$.
    In \cref{fig:pno_conv}c, we show the convergence of the DLPNO-RPA reaction energy for forming androstenedione and pyrocatechol from the precursor molecule shown in \cref{fig:pno_conv}a.
    A corresponding plot for the binding energy of the coronene dimer is presented in \cref{fig:pno_conv}d.
    In both cases, the error introduced by the DLPNO approximation is well below $0.5$~kcal/mol for $T_{\textrm{PNO}} \leq 3\times10^{-7}$.
    Consistent with previous studies~\cite{Neese09JCPa}, applying the PNO correction (\ref{eq:pno_correction}) is crucial for achieving this level of accuracy.
    The results without the PNO correction, also shown in \cref{fig:pno_conv}, highlight its importance, as the uncorrected values fall short of the desired precision unless a much tighter $T_{\textrm{PNO}} \leq 10^{-8}$ is used.
    The accuracy and convergence pattern are very similar for DLPNO-RPA+SOSEX, as shown in FIG.~S1 in the Supporting Information.

    \begin{figure}[!b]
        \centering
        \includegraphics[width=8cm]{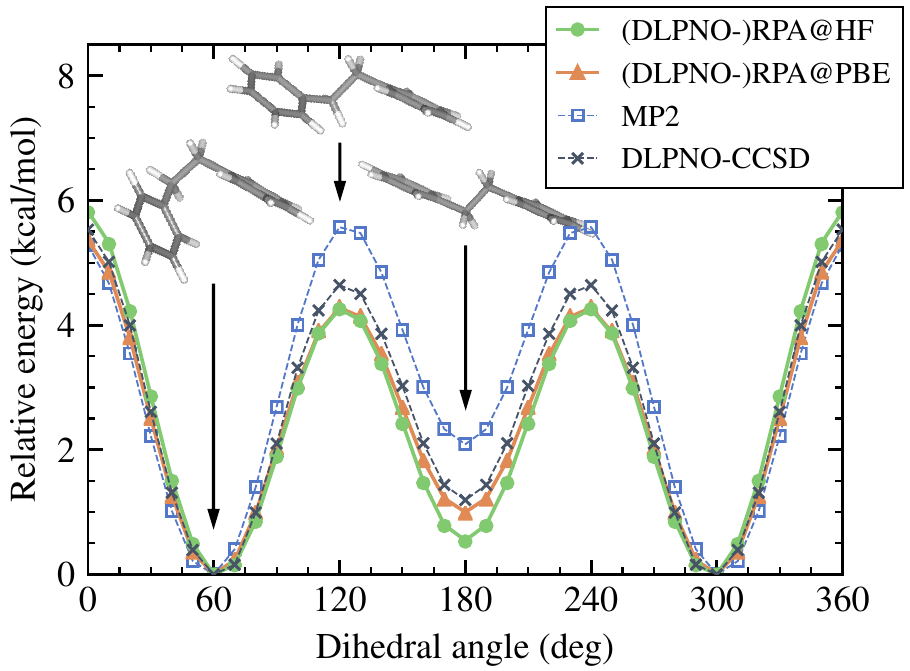}
        \caption{Potential energy surface of the pedal motion in a bibenzyl molecule predicted by DLPNO-RPA using both the HF (green circles) and PBE (orange triangles) references.
        The canonical RPA@HF and RPA@PBE results are denoted by lines of the corresponding color.
        MP2 and DLPNO-CCSD results are also included for comparison.
        All calculations were performed using the def2-TZVP/def2-TZVP-RI basis sets with fixed geometries obtained from a relaxed scan of the dihedral angle at B3LYP+D3(BJ)/def2-TZVP level.
        }
        \label{fig:pes}
    \end{figure}

    In \cref{fig:pes}, we demonstrate the ability of DLPNO-RPA to generate accurate and smooth potential energy surfaces by examining the dihedral rotation in a bibenzyl molecule.
    This molecule serves as a prototype for the pedal motion frequently observed in covalent organic framework materials~\cite{Chi23NC,Frimpong24PCCP}.
    Using the default parameters listed in \cref{tab:param}, DLPNO-RPA achieves quantitative accuracy compared to canonical ACFDT-RPA, producing a potential energy curve that aligns perfectly with the latter.
    This holds true for both HF and PBE mean-field references.
    In \cref{fig:pes}, we also include results from MP2 and DLPNO-CCSD (obtained using ORCA 5.0~\cite{Neese20JCP}) for comparison.
    DLPNO-RPA offers a significant improvement over MP2 by accurately capturing the relative heights of the barriers at $0$ and $120$ degrees and providing a better description of the well depth at $180$ degrees.
    In addition, DLPNO-RPA with a PBE reference consistently outperforms the HF reference, particularly in describing the well depth at $180$ degrees.

    \begin{figure}[!t]
        \centering
        \includegraphics[width=8cm]{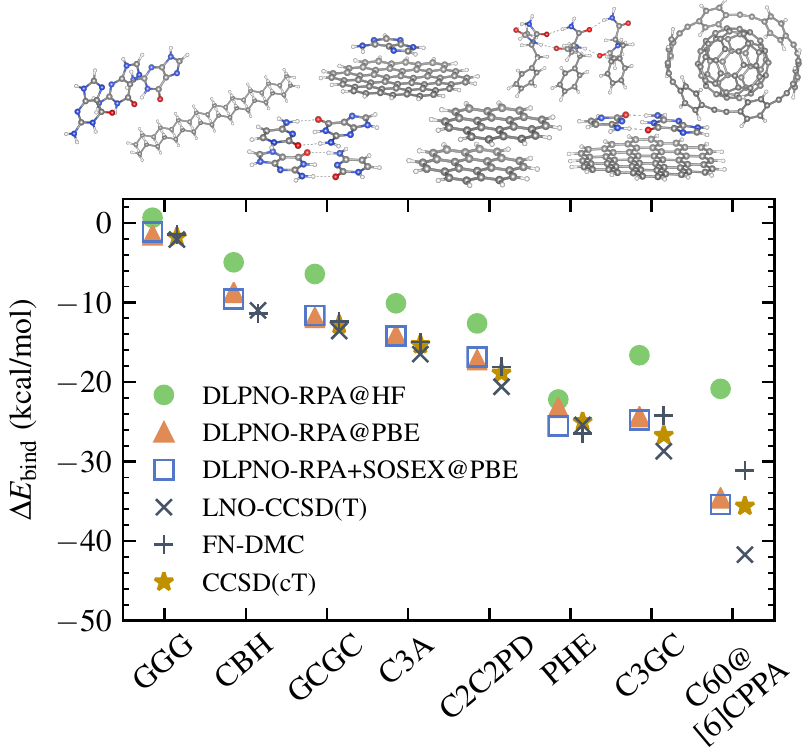}
        \caption{Binding energies of eight large molecules (the L7 set~\cite{Sedlak13JCTC} and \ce{C_{60}}@[6]CPPA) predicted by DLPNO-RPA using both the HF and the PBE mean-field references.
        The results have been extrapolated to the CBS limit as noted in the main text.
        For the PBE reference, we also include results from DLPNO-RPA+SOSEX.
        The LNO-CCSD(T) and FN-DMC reference results were taken from ref~\cite{AlHamdani21NC}.
        The CCSD(cT) reference results were taken from ref~\cite{Schafer24arXiv}.
        }
        \label{fig:L8}
    \end{figure}

    One attractive feature of RPA is its ability to describe weak interactions important for applications involving molecular crystals~\cite{Lu09PRL,Klimes16JCP,Pham23JCP}, porous materials~\cite{Chehaibou19JCTC}, and solid surfaces~\cite{Schimka10NM,Ren09PRB,Olsen11PRL}.
    We demonstrate this by calculating the DLPNO-RPA binding energies for a set of eight molecules, including the L7 set~\cite{Sedlak13JCTC} and the \ce{C60}@[6]CPPA complex, as shown in the top of \cref{fig:L8}.
    Highly accurate reference binding energies were reported for this set of molecules in a previous study~\cite{AlHamdani21NC} using CCSD(T) (with the local natural orbital approximation~\cite{Nagy17JCP,Nagy19JCTC}) and fixed-node diffusion Monte Carlo~\cite{Foulkes01RMP} (FN-DMC).

    We first verified using the HF reference and the cc-pVTZ/cc-pVTZ-RI basis sets that DLPNO-RPA achieves high accuracy in the predicted binding energies across the entire set, with a mean absolute error (MAE) of $0.27$~kcal/mol compared to canonical ACFDT-RPA (FIG.~S2 in the Supporting Information).
    We then extrapolated the DLPNO-RPA results to the CBS limit using calculations with TZ and QZ basis sets.
    The DLPNO approximation is essential for this extrapolation, as canonical RPA calculations with QZ basis sets become prohibitively expensive for the largest molecules in the set.
    The final DLPNO-RPA binding energies, obtained using both HF and PBE references, are presented in \cref{fig:L8}.
    Compared to the FN-DMC reference, DLPNO-RPA@HF consistently underestimates the binding energy, resulting in a MAE of $5.9$~kcal/mol.
    The situation is significantly improved by using a PBE reference, with DLPNO-RPA@PBE reducing the MAE to $1.5$~kcal/mol, approaching chemical accuracy.
    The SOSEX correction has a minor effect for all dispersion-bound molecules, but shows improvement for the phenylalanine trimer (PHE), where the binding energy, dominated by hydrogen bonding, is particularly sensitive to the treatment of exchange~\cite{Ren13PRB,Chedid21JCP,Tahir24JPCA}.

    In ref~\onlinecite{AlHamdani21NC}, a significant discrepancy was noted between the FN-DMC and CCSD(T) binding energies for \ce{C60}@[6]CPPA, the largest molecule in the set.
    Sch\"{a}fer and co-workers~\cite{Schafer24arXiv} attributed this difference to the systematic overestimation of interaction energies by finite-order perturbation theories when dealing with highly polarizable systems~\cite{Nguyen20JCTC} such as molecular crystals~\cite{Liang23JPCL}.
    This observation aligns with the significantly overestimated binding energies predicted by MP2 for these molecules (FIG.~S3 in the Supporting Information), which further explains the favorable performance of (DLPNO-)RPA reported above, which resums the direct ring diagram present in MP2 to infinite order.
    To address the issue of CCSD(T), Sch\"{a}fer and co-workers employed the recently developed CCSD(cT) method~\cite{Masios23PRL}, which improves upon CCSD(T) by incorporating selected higher-order terms in the triple excitation approximation.
    The resulting CCSD(cT) binding energies, also shown in \cref{fig:L8}, show better agreement with FN-DMC.
    Notably, our DLPNO-RPA@PBE binding energies follow closely the CCSD(cT) results across the entire set, including the challenging case of \ce{C60}@[6]CPPA.
    This comparison underscores the potential of DLPNO-RPA as a cost-effective alternative to higher-order coupled cluster methods for accurately describing weak interactions.


    \begin{figure}[!t]
        \centering
        \includegraphics[width=8cm]{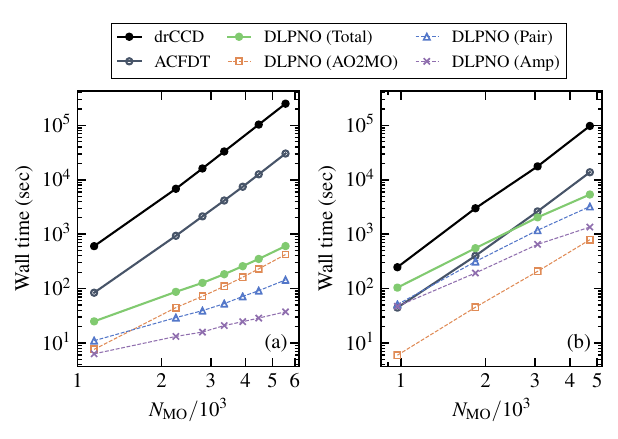}
        \caption{Wall time for DLPNO-RPA calculations with $T_{\textrm{PNO}} = 3\times10^{-7}$ using the cc-pVTZ/cc-pVTZ-RI basis sets for (a) glycine peptide chains and (b) diamond clusters of increasing size.
        The breakdown of wall time for different components of the DLPNO-RPA calculations is shown, along with canonical RPA results obtained using both drCCD (with the DF-based quartic-scaling algorithm~\cite{Kallay15JCP}) and ACFDT for comparison.
        The cost of HF is not included.
        All calculations were performed on a single node using 24 AMD EPYC 9474F CPU cores and 8 GB of memory per core, utilizing OpenMP for parallelization.
        }
        \label{fig:cost}
    \end{figure}

    Finally, we illustrate the computational efficiency achieved with the DLPNO approximation.
    In \cref{fig:cost}, we compare the wall time of DLPNO-RPA to canonical RPA obtained using drCCD (with the DF-based quartic-scaling algorithm~\cite{Kallay15JCP}) and ACFDT for a series of glycine chains (\cref{fig:cost}a) and carbon quantum dots (CQDs) (\cref{fig:cost}b) of increasing size.
    The DLPNO approximation proves highly effective for glycine chains, where a $3$-fold/$25$-fold speedup over canonical ACFDT/drCCD-RPA is observed as early as \ce{(Gly)_{8}} (approximately $1,000$ orbitals), increasing to a $50$-fold/$400$-fold speedup for \ce{(Gly)_{40}} (around $5,600$ orbitals).
    For CQDs, DLPNO-RPA is still more efficient than the canonical drCCD-RPA, where a two-fold speedup is observed for a small CQD \ce{C_{26}H_{32}} (approximately $1,000$ orbitals), increasing to a $20$-fold speedup for the largest CQD \ce{C_{136}H_{104}} (approximately $4,700$ orbitals).
    In contrast, the crossover between DLPNO-RPA and canonical ACFDT-RPA for CQDs occurs at a larger system size of about $2,500$ orbitals, and a modest three-fold speedup is achieved for the largest CQD studied.
    These results are noteworthy, especially given that canonical ACFDT-RPA calculates only the correlation energy, whereas DLPNO-RPA also computes the wavefunction amplitudes.

    To provide deeper insights, \cref{fig:cost} also presents a detailed breakdown of the DLPNO wall time, separating the contributions from the integral transform (\ref{eq:irP}), pair ERI assembly (\ref{eq:pao_eris}), and the solution of the amplitude equations (\ref{eq:rpa_amp_eqn}).
    Additionally, FIG.~S4 and S5 in the Supporting Information include data on domain sizes and pair counts to further elucidate these trends.
    For glycine chains, the one-dimensional structure leads to a rapid saturation of the domain size and efficient pair screening.
    As a result, the pair ERI assembly and the solution of amplitude equations both exhibit linear scaling with a small prefactor, making the cubic-scaling integral transform the primary computational cost.
    In contrast, the three-dimensional nature of CQDs results in a slower saturation of domain size and a superlinear increase in both strong and weak pair counts, even for the largest system considered.
    Consequently, the overall DLPNO-RPA cost for CQDs is dominated by calculating pair ERIs and solving amplitude equations, both of which display quadratic scaling within the tested system size range.


    \section{Concluding remarks}

    In summary, we have developed DLPNO-RPA to enable efficient reduced-scaling RPA calculations for large molecular systems.
    As byproducts, this work has also enhanced the local correlation infrastructure of PySCF, producing its first implementations of domain constructions, local DF, and local MP2 (both OSV-MP2~\cite{Yang11JCP} and DLPNO-MP2~\cite{Pinski15JCP}).
    Our results show that DLPNO-RPA consistently achieves sub-kcal/mol accuracy in reaction energies, potential energy surfaces, and weak interactions compared to canonical ACFDT-RPA, while significantly reducing computational costs depending on the dimension of the system.
    As a practical application, we demonstrated that highly accurate binding energies in the CBS limit can be efficiently calculated using DLPNO-RPA for large molecules, and the results exhibit excellent agreement with those obtained from other high-level correlated wavefunction methods.
    These findings pave the way for the routine application of RPA-based approaches, exemplified by DFT with double-hybrid functionals~\cite{Grimme16PCCP}, in molecular quantum chemistry.

    \section*{Supporting Information}

    See Supporting Information for (i) molecular structures, (ii) PNO convergence for DLPNO-RPA+SOSEX, (iii) binding energy of DLPNO-RPA in TZ basis sets, (iv) binding energy of MP2, and (v) domain and pair counts for glycines and CQDs.

    \section*{Conflict of interest}
    The authors declare no competing conflicts of interest.

    \section*{Acknowledgments}

    This work was supported by the National Science Foundation under Grant No.~CHE-1848369 (T.C.B.~and H.-Z.Y.), the Columbia Center for Computational Electrochemistry (Y.H.L.~and H.-Z.Y.), startup funds from the University of Maryland, College Park (H.-Z.Y.), and US Department of Energy, Office of Science, Award No.~DE-SC0023318 (X.Z.~and G.K.L.C.).
    We also acknowledge computing resources provided by the Flatiron Institute and the Division of Information Technology at the University of Maryland, College Park.
    The Flatiron Institute is a division of the Simons Foundation.

    \bibliography{refs}

\end{document}


\title{Supporting Information for Efficient Implementation of the Random Phase Approximation with Domain-based Local Pair Natural Orbitals}

    \author{Yu Hsuan Liang}
    \affiliation{Department of Chemistry, Columbia University, New York, NY, 10027, USA}
    \author{Xing Zhang}
    \affiliation{Division of Chemistry and Chemical Engineering, California Institute of Technology, Pasadena, California 91125, USA}
    \author{Garnet Kin-Lic Chan}
    \affiliation{Division of Chemistry and Chemical Engineering, California Institute of Technology, Pasadena, California 91125, USA}
    \author{Timothy C. Berkelbach}
    \email{tim.berkelbach@gmail.com}
    \affiliation{Department of Chemistry, Columbia University, New York, NY, 10027, USA}
    \affiliation{Initiative for Computational Catalysis, Flatiron Institute, New York, NY, 10010, USA}
    \author{Hong-Zhou Ye}
    \email{hzye@umd.edu}
    \affiliation{Department of Chemistry and Biochemistry, University of Maryland, College Park, MD, 20742, USA}
    \affiliation{Institute for Physical Science and Technology, University of Maryland, College Park, MD, 20742, USA}

    \maketitle

    \vspace{3em}

    Further supplementary data can be found in the following Github repository
    \begin{center}
        \url{https://github.com/hongzhouye/supporting_data/tree/main/2024/DLPNO_RPA}
    \end{center}


    \section{Geometries}

    \begin{itemize}
        \item Structures for androstenedione precursor and the decomposition products pyrocatechol and androstenedione are taken from ref~\onlinecite{Werner15JCTC}.

        \item Structures for the biphenyl molecule are obtained by a relaxed scan at B3LYP+D3(BJ)/def2-TZVP level using ORCA~\cite{Neese20JCP}.
        The XYZ files are available in the Github repository.

        \item Structures for the L7 set are taken from ref~\onlinecite{Sedlak13JCTC} directly, while that of \ce{C_{60}}@[6]CPPA is taken from ref~\onlinecite{AlHamdani21NC}.

        \item Structures for the glycine chains are taken from ref~\onlinecite{Werner15JCTC}.

        \item Structures for the carbon quantum dots are carved out from diamond crystal, saturating the surface with hydrogen atoms, and then optimized at B3LYP/def2-SVP level.
        The XYZ files are available in the Github repository.
    \end{itemize}

    %

    \section{Supplementary figures}

    \begin{figure}[!h]
        \centering
        \includegraphics[width=12cm]{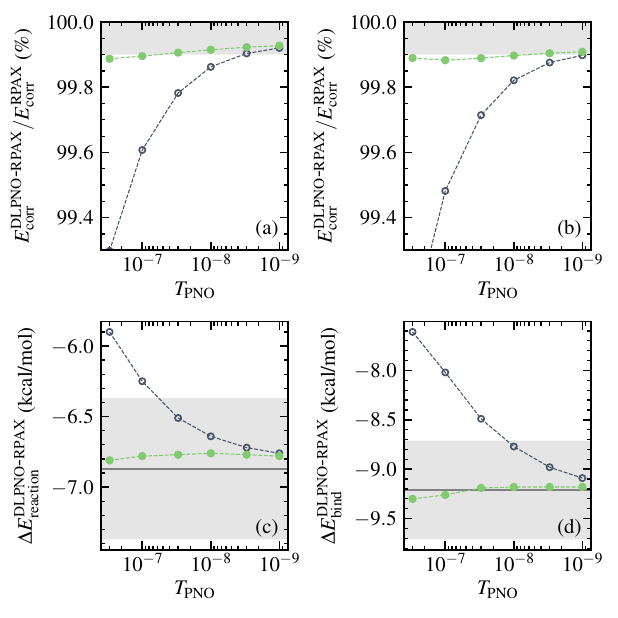}
        \caption{A plot similar to Fig.~\fakeref{M1} showing the convergence of DLPNO-RPA+SOSEX (denoted by ``RPAX'' in the plot).
        The error is calculated against canonical RPA+SOSEX, obtained using the quartic scaling CC-RPA~\cite{Kallay15JCP} implemented in a developer version of PySCF.}
        \label{fig:pno_conv}
    \end{figure}

    \begin{figure}[!h]
        \centering
        \includegraphics[width=9cm]{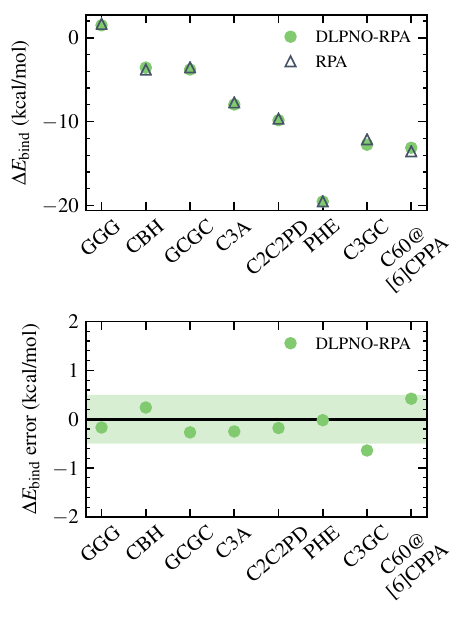}
        \caption{A plot similar to Fig.~\fakeref{M2} showing the binding energy calculated by DLPNO-RPA compared to canonical ACFDT-RPA (top) and the error (bottom) using a HF reference and cc-pVTZ/cc-pVTZ-RI basis sets.
        The shaded area in the error plot indicates error less than $\pm 0.5$~kcal/mol.}
        \label{fig:L8_tz}
    \end{figure}

    \begin{figure}[!h]
        \centering
        \includegraphics[width=9cm]{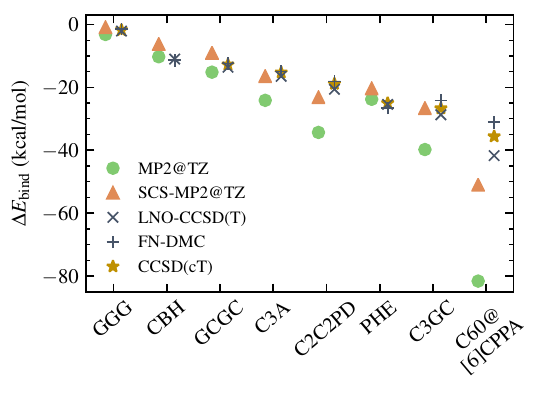}
        \caption{A plot similar to Fig.~\fakeref{M2} shows the binding energy calculated by MP2 and SCS-MP2, compared to reference values from coupled-cluster and diffusion Monte Carlo methods discussed in the main text.
        The (SCS-)MP2 calculations were performed using cc-pVTZ/cc-pVTZ-RI basis sets.
        The residual basis set incompleteness errors are estimated to be a few kcal/mol, based on the difference between DLPNO-RPA@TZ and DLPNO-RPA@CBS.
        These residual errors are positive, indicating that using larger basis sets would yield binding energies of even greater magnitude, thus amplifying the overbinding trend observed in MP2.
        }
        \label{fig:L8_mp2}
    \end{figure}

    \begin{figure}[!h]
        \centering
        \includegraphics[width=12cm]{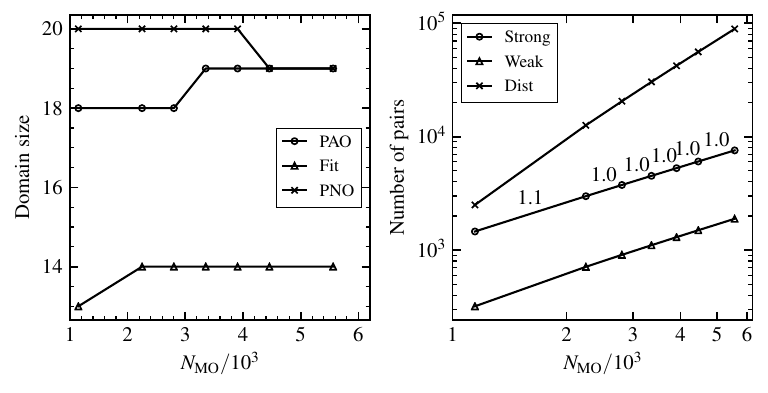}
        \caption{(Left) Average size of PAO, fitting, and PNO domains for glycine chains as a function of molecular size, indicated by the number of orbitals.
        For PAO and fitting domains, the average number of atoms is shown, while for the PNO domain, the average number of PNOs is presented.
        Averages are rounded to the nearest integer.
        (Right) Number of strong, weak, and distant pairs as a function of molecular size.
        The number of strong pairs increases linearly with molecular size, even for the smallest molecule tested, as indicated by the scalings shown in the figure, which were obtained through a series of two-point fits.
        }
        \label{fig:glyn_domain}
    \end{figure}

    \begin{figure}[!h]
        \centering
        \includegraphics[width=12cm]{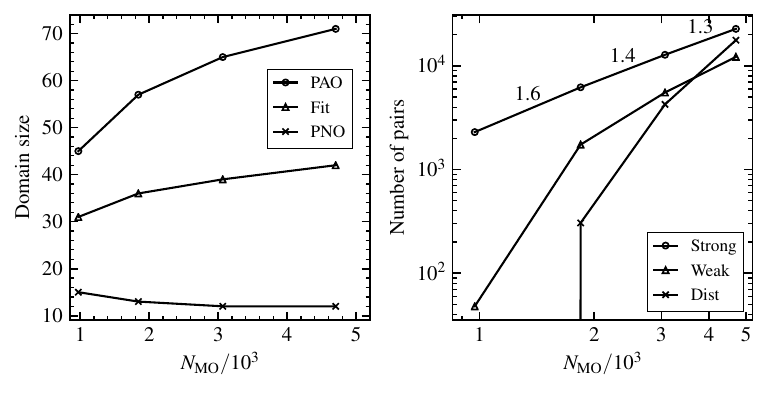}
        \caption{Plots similar to \cref{fig:glyn_domain} for the carbon quantum dots.
        The scalings for the number of strong pairs shown in the right panel suggest that linear scaling is not achieved even for the largest molecule tested.
        }
        \label{fig:diamond_domain}
    \end{figure}

    \clearpage

    \bibliography{refs}